\documentclass{ceab}   

\usepackage{epsfig}     
\usepackage{graphicx}   
\usepackage{url}
\usepackage{ceabbib}     
\usepackage[T1]{fontenc}

\begin{document}

\def\tit{The Gaia mission}
\def\aut{Eyer, Holl, Pourbaix et al.}

\title{The Gaia mission}

\author{L. Eyer$^{1}$, B. Holl$^{1}$, D. Pourbaix$^{2, 3}$, N. Mowlavi$^{1}$, 
C. Siopis$^2$, F. Barblan$^{1}$, \\
D. W. Evans$^{4}$ and P. North$^{5}$
\vspace{2mm}\\
\it $^1$Observatoire de Gen\`{e}ve, D\'{e}partement d'Astronomie, \\ 
\it     Universit\'{e} de Gen\`{e}ve, 1290 Sauverny, Switzerland\\
\it $^2$Institut d'Astronomie et d'Astrophysique, \\
\it     Universit\'{e} Libre de Bruxelles, 1050 Bruxelles, Belgium\\ 
\it $^3$Senior Research Associate, FNRS, 1000 Bruxelles, Belgium\\ 
\it $^4$Institute of Astronomy, University of Cambridge,\\
\it     Cambridge, CB3 0HA, United Kingdom\\
\it $^5$Laboratoire d'astrophysique, Ecole Polytechnique F\'{e}d\'{e}rale de Lausanne (EPFL), \\
\it     Observatoire de Sauverny, 1290 Versoix, Switzerland 
}

\maketitle

\begin{abstract}

Gaia is a very ambitious mission of the European Space Agency. At the heart of Gaia lie the measurements of the positions, distances, space motions, brightnesses and astrophysical parameters of stars, which represent fundamental pillars of modern astronomical knowledge. We provide a brief description of the Gaia mission with an emphasis on binary stars. In particular, we summarize results of simulations, which  estimate the number of binary stars to be processed to several tens of millions. We also report on the catalogue release scenarios. In the current proposal, the first results for binary stars will be available in 2017 (for a launch in 2013). 

\end{abstract}


\section{The Gaia mission}

Gaia is a cornerstone mission of the European Space Agency. It will perform an all-sky survey and observe all objects brighter than magnitude $V \sim $ 20. With this constraint, the number of objects observed will reach more than 1 billion. Gaia observations consist of astrometric, photometric and spectroscopic measurements, leading to an unprecedented description of the Galaxy and its components. The multi-epoch nature of the mission will moreover allow the systematic detection, characterization and
classification of all sorts of variable objects. The spacecraft will be launched in 2013 by a Soyuz-Fregat rocket from French Guiana. The duration of the mission is fixed to five years, with a possible extension of one year. The final results are expected to be released in 2021-2022.

The Gaia Focal Plane is covered by 106 CCDs forming a camera of nearly one billion pixels. There are three main instruments:
\begin{itemize}
\item The astrometric field (AF), recording white light G-band (over 330-1050 nm), and from which the astrometry will be derived.
\item The spectro-photometry, composed of a Blue Photometer (BP) covering the wavelength range from 330 to 680~nm, and a Red Photometer (RP) covering 640 to 1050~nm. Both
instruments produce low resolution spectra consisting of 62 pixels from which an integrated magnitude will be computed.
\item The Radial Velocity Spectrometer (RVS), a near-infrared instrument with a resolution of 11,500, that spans the wavelength range from 847 to 871~nm\footnote{the
wavelength range has been truncated by 3~nm with respect to the initial requirements due to a required replacement of the RVS filter}. This range covers the Calcium
triplet. Data from the RVS will be available for stars brighter than $V \sim 16$~mag.
\end{itemize}

Gaia has two telescopes with a field of view of 0.45 deg$^2$ each, separated by 106.5~deg and rotating at a rate of 60 arc sec sec$^{-1}$.
Due to its motion, described by the scanning law, Gaia scans the sky in precessing great circles. This causes each source to be observed 
for on average 72 field-of-view transits over the 5 year mission.

\subsection{Expected Performance}

The end-of-mission performance of the spacecraft can be found on the Gaia page of the ESA website (\url{http://www.rssd.esa.int/Gaia}), under ``Science Performance''. They are consistent with the initial requirements, with only few minor non-compliances.

\begin{figure}
  \begin{center}
       \epsfig{file=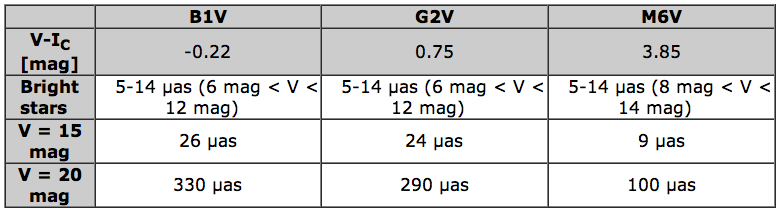,width=12cm}%
  \end{center}
  \caption{\label{fig:astroperf} Astrometric performance summary as found on the Gaia website (January 2013).}
\end{figure}

The astrometric performance, given in terms of the mean parallax error at the end of the mission, depends on the magnitude of the considered object and its colour (see Fig.~\ref{fig:astroperf}).

On average, the errors on the position and the proper motion of the stars at the end of the mission can be derived from their parallax error by multiplying the latter by a factor
of 0.7 and 0.5, respectively.

The end-of-mission errors on the photometry in the G, integrated BP and RP bands, and on the radial velocity derived from the RVS instrument, are also given on the ESA website cited above.
The per transit photometric performance, on the other hand, has been presented in Eyer, Palaversa, Mowlavi, et al. (2012). The photometric precision in the G-band from the astrometric field may reach the milli-magnitude level at magnitudes brighter than 14~mag. It is worth mentioning that this photometric precision is still estimated to be 20~mmag at the limiting magnitude of 20~mag. The survey is actually not limited in magnitude by the signal-to-noise ratio but by the limited data transfer rate from the L2 Lagrangian point (location of the spacecraft) to the Earth.
A brief comparison of the astrometric and photometric performances between Gaia and LSST is presented in Eyer, Dubath, Mowlavi, et al. (2012).

\section{The Data Processing and Analysis Consortium}

The role of the DPAC consortium is to produce a Gaia catalogue containing useful quantities for the scientific exploitation of the mission. The consortium currently includes more than 400 active members, and is divided into Coordination Units (CUs) that take care of thematic activities. As examples of themes, we may cite: the astrometric reduction, which produces the five astrometric parameters (position, proper motion and parallax) for every object; the photometric reduction, which produces calibrated photometry and spectro-photometry; variability characterization, which characterizes and classifies variable objects, among which periodic variable stars, etc\ldots
The CUs develop mathematical methods and algorithms related to their topic, and each of them is associated with a Data Processing Centre (DPC) that has the responsibility to process the Gaia data as it becomes available.

In the next two sections, we give more details on two CUs, one dedicated to the analysis of variable objects, and another dedicated to objects processing, which takes care in particular of the astrometry of non-single stars. For more details, the reader is referred to Mignard et al. (2008).

\subsection{Gaia Variability Processing and analysis}

Coordination Unit 7 (CU7) is responsible for the data processing and analysis of the variable sources observed by Gaia. CU7 will study the photometric and spectroscopic time series produced by the spacecraft after they have been calibrated by two other coordination units (CU5, CU6).

The number of Gaia sources expected to be variable is estimated to range between 50 and 150~million. Processing such a large number of variable sources, that moreover include a variety of extremely diverse variability behaviours, is a challenge. To meet this challenge, about 70 people from a dozen institutes have come together to constitute CU7. The effort is coordinated by the University of Geneva.

The main tasks of CU7 are as defined in the ESA Announcement of Opportunity:
(a)~variability classification, period search and variability models; 
(b)~feedback of variability analysis on the calibration models;
(c)~catalogue exploration and checks of the variability database; 
(d)~statistical analysis of the data; 
(e)~external observation coordination; 
(f)~variability announcements.

\subsection{Gaia Object Processing and Analysis}

Coordination Unit~4 (CU4) is responsible for Objects Processing: its tasks include the processing of the astrometric and photometric data of more complex objects not handled by the astrometric core processing, and specifically: 
(a)~non-single stars (binary and multiple stars);
(b)~Solar System objects (asteroids, near-Earth objects, etc);
(c)~extended objects.

CU4 gathers about 50 scientists from seven countries and will rely upon the French Centre National d'Etudes Spatiales (CNES) computing facilities for its processing.
This effort is coordinated by the University of Brussels (ULB).

\subsection{Gaia data releases}

Data processing within the Gaia consortium will proceed iteratively, based on cycles of, typically, 6 months or 1 year. The Gaia Science Team, in coordination with the DPAC
consortium Executive, agreed on a scenario for the data releases of the mission based on this cyclic development of the data products (cf. O'Mullane \& van Leeuwen 2012, and Prusti 2012). 
As of today (January 2013), five releases are foreseen, with the following timeline (we mark in bold the releases of direct concern to binary stars):

\begin{itemize}
\item First release (launch + 22 months): 
It will contain the positions and the mean G magnitudes of apparently single non-variable sources, together with their associated errors. In addition, a catalogue of proper motions will be provided by combining the Gaia positions with those of Hipparcos, named the Hundred Thousand Proper Motions (HTPM) catalogue.
\item Second release (launch + 28 months): As the length of accumulated data will allow to decouple parallax and proper motion, the release will contain the five parameters of the astrometric solution for each source, i.e., the position, proper motion and parallax, together with the associated errors. 
This release will concern only apparently single sources. In the photometry, there will be the mean G magnitude and the first results of the mean integrated BP/RP spectro-photometry for which astrophysical parameter estimations have been verified.
The first results of the RVS instrument, including the mean radial velocity, will be provided for stars which show non-variable behaviour.
\item Third release (launch + 40 months):
It will contain an update of all the quantities published in the previous releases.
In addition, it will contain {\bf astrometric orbital solutions for non-single stars for periods between 2 months and 75\% of the observation duration}, as well as mean spectro-photometry BP, RP, RVS products for stars with their associated astrophysical parameters.
\item Fourth release (launch + 65 months):
In addition to the above, a catalogue of the astrophysical parameters of the stars will be provided, together with the mean RVS spectra. The first global {\bf Variable Star catalogue} will be presented, as well as the {\bf non-single star catalogue} and Solar System results.
\item Fifth release (end-of-mission operation + 3 years): This is the final release. It will contain all the data products that have been defined:  astrometry, photometry, spectra, radial velocities, {\bf products on non-single stars}, {\bf list of exoplanet candidates}, solar system objects, the \bf{variable star catalogue} and astrophysical parameters.

\end{itemize}

As there are other photometric surveys making global analyses of variability, it is important that the results of the Gaia variability analysis are made available to researchers as soon as possible.  CU7 should prepare releases, which contain the data of given variability types as soon as the errors on their classification, basically affected by contamination and (in)completeness, are adequately estimated. The sub-catalogues of short-period and large-amplitude variable stars, such as RR~Lyrae stars, are therefore likely to be the first ones to be released.

The data analysis done by the consortium may require auxiliary data such as literature data and/or dedicated ground-based observations. Where this is the case, these data will also be made public.

Finally, it must be mentioned that there is no proprietary period for the scientific analysis of the data.

\subsection{Gaia alerts}

In addition to the above releases, alerts will be issued as necessary. An alert should be activated whenever science return would be lost if no quick ground-based follow-up is performed. The photometric detection of such cases and the actual delivery of these alerts are under the responsibility of  the Institute of Astronomy of the University of Cambridge. Astrometric alerts will be issued for solar system objects. These latter alerts will be issued by CU4 with the help of the IAU Minor Planet Centre (operated by the Smithsonian Astrophysical Observatory).

Cataclysmic and eruptive phenomena, often due to interacting binaries, are examples of transients that may require alerts when detected. Detection, classification and rapid response are crucial steps in order that these transient objects be followed up and exploited scientifically. The process from on-board data acquisition to ground-based first calibration is not immediate, though. The possible reaction time ranges from a couple of hours to 48~hours.

A ground-based verification phase (estimated to last 3 months) will start after Gaia enters in operation.  So far, some tests have been performed on transients detected by currently operating surveys that were followed up by chosen ground-based telescopes. For example, a few transients alerted by the Catalina survey have been followed up by the 1.2~m Euler Telescope (Mahabal et al.~2012).

\section{Binary stars of the Gaia mission}

\begin{figure}
\begin{center}
 \epsfig{file=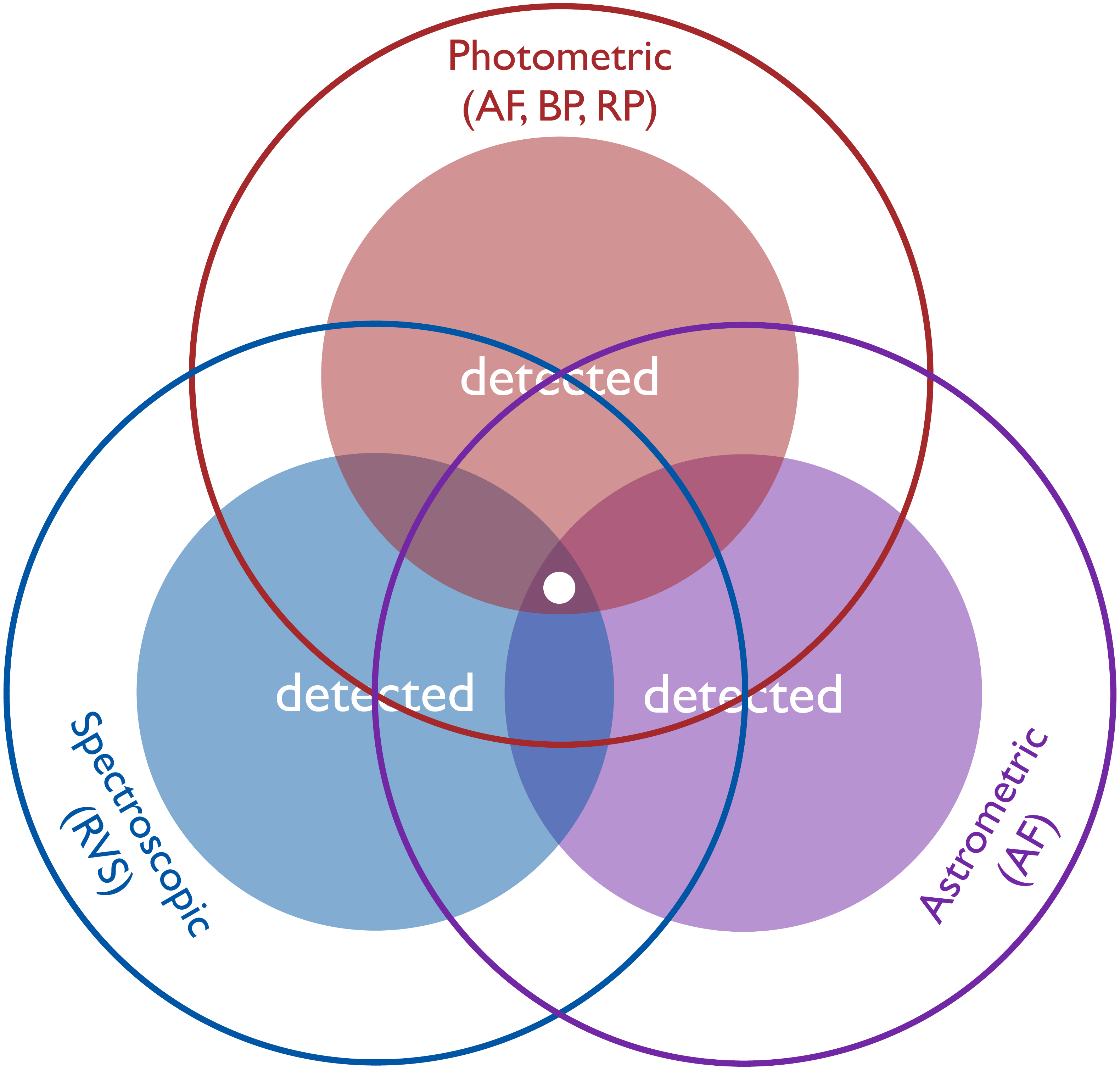,width=10cm}%
\end{center}
\caption{\label{fig:venndigram} Venn diagram of the binary signal to which the different Gaia instruments are sensitive. The signal of astrometric binaries can be detected from the centroid position 
of the AF observations. The signal of spectroscopic binaries can be detected from the spectra provided by the RVS instrument. The signal of photometric binaries (mainly eclipsing binaries and ellipsoidal binaries) can be detected by photometric variations in the AF, BP and RP instruments. The white spot at the centre represents the non-detection of binaries which have all 3 signals.}
\end{figure}

The study of binary stars will greatly benefit from the Gaia mission, with significant contributions from each of the astrometry, photometry and RVS instruments on board (see Fig.~\ref{fig:venndigram}).
Within the consortium, the processing of astrometric binaries is done by CU4, that of spectroscopic binaries by CU6, CU4 and CU7, and that of eclipsing binaries by CU7 and CU4. Binaries are probably the objects which lead to the most interrelations within Gaia measurements and within the consortium. That makes them difficult to process, but also scientifically very valuable.

The large number of binaries detected will allow to get statistical descriptions and global properties of populations of binary stars. The fact that Gaia is an all-sky survey with data homogeneously obtained by its instruments allows to compare different stellar populations more easily. For example, the binary fraction and the orbital properties of binaries will be determined, and their dependence on factors like metallicity can be investigated.

In addition, the Gaia survey will allow to discover very rare and unusual binaries. Finding these rare cases is another challenge that the consortium has to face. The consortium should thus provide a correct high-level classification and good descriptive parameters to help such scientific searches. Examples of particularly interesting cases include catching objects in a rapid phase of binary evolution, testing stellar evolution models (e.g. Torres et al.~2010),  finding extreme masses (Bonanos et al.~2004, Rauw et al.~2004) or extreme orbital properties like very short periods (Nelemans~2005) or large apsidal motions (North et al. 2010), or calibrating eclipsing binaries as distance indicators (Rucinski~1996, Paczynski~1997) with their parallax.

Cases of binaries where one component is a pulsating star are also particularly interesting. The presence of a pulsating star in the system may allow the use of the Baade-Wesselink method or asteroseismology. Comparing the astrophysical parameters obtained from these two independent methods (derived from binary and pulsation) would then be most fruitful.

\subsection{Expected harvest from Gaia}
In order to better pin-down the size of the processing effort and to prepare the Gaia data analysis, a data set of about 1\% of stars of the Galaxy that will be observed by Gaia has been simulated by CU2, the Coordination Unit dedicated to produce simulations. 
These simulations are based on a model of the Galaxy (see Robin et al.~2012) and on models of binary stars (Arenou 2011). In the past, there have  been several attempts to estimate the numbers of various classes of objects that Gaia will detect. For example, the estimated number of eclipsing binary detections has ranged from half a million to 7 million (see Eyer, Dubath, Mowlavi, et al.~2012 and references therein). The large variations in these estimates show how our knowledge is somehow limited. 
We can also estimate the number of various objects using the above-mentioned simulations. Since 10.5 million sources have been simulated, we apply a corrective factor of 100 to reach the 1 billion stars that Gaia is expected to observe. Doing this, we find that 30 million objects will be processed as astrometric non-single stars, 8 million as spectroscopic binaries (of which 59\% will be SB2) and 4 million as eclipsing binaries (of which 12\% will be spectroscopic binaries).


\subsection{Processing of eclipsing binaries}
\begin{figure}
\begin{center}
 \epsfig{file=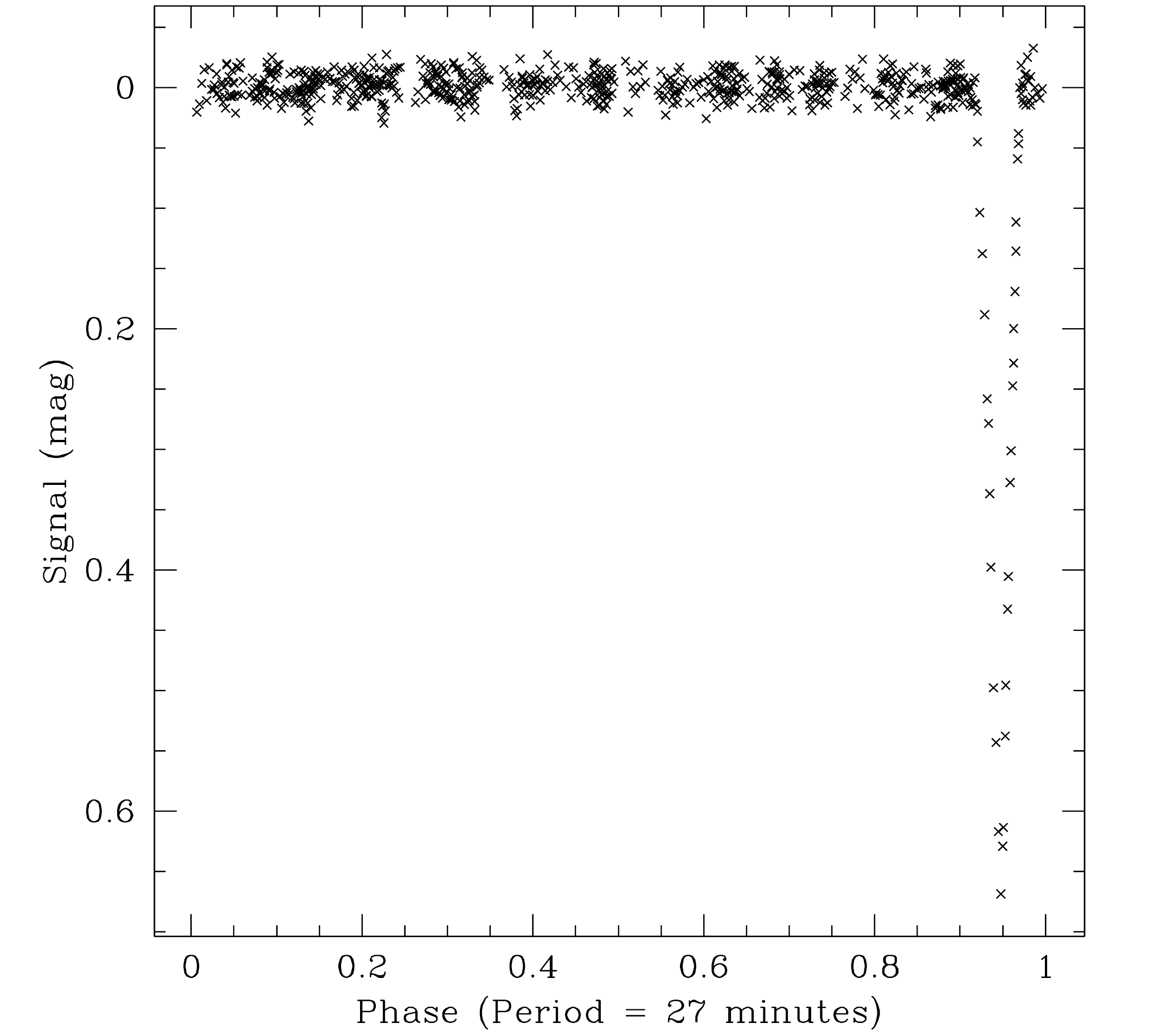,width=8cm}%
\end{center}
\caption{\label{fig:AMCVn} A simulation of an eclipsing AM CVn star of 27 minute period as measured by Gaia with the per-CCD resolution.}
\end{figure}

We finish this document with the example of eclipsing binaries, which are particularly difficult to process and which require the most intense interactions between different Coordination Units. The different steps required to process these objects are the following. First, astrometry, photometry and spectroscopy calibrations have to be done. Then, the sources are processed in the CU7 pipeline, on a per-source basis, with four sequential main tasks:
\begin{itemize}

\item (1)~{\bf Special Variability Detection}. This module contains general statistical tests to detect variability, as well as special statistical tests to identify extra variable stars that may not have been detected as such with the general statistical tests. Those special tests take advantage of known properties of particular types of variable stars, such as planetary transits, (magnetic) solar-like variable stars, or short period objects (less than 2 hours) such as eclipsing AM CVn binary stars. Gaia will provide a systematic all-sky search of all these objects.
Figure~\ref{fig:AMCVn} shows an example of an AM CVn star with simulated Gaia data.  The special tests are made possible thanks to the 9 AF CCD observations that are available for each field-of-view transit and which are separated by 4.4~seconds. 
Furthermore, the irregular sampling of Gaia allows to detect very high frequencies (Eyer \& Bartholdi 1999). 
 
\item (2)~{\bf Characterization} of this variability. It contains statistical parameters describing the variability, such as moments of the distribution, periods, and light-curve model parameters. Several period search methods have been implemented, among which the standard Deeming, Lomb-Scargle, and generalized Lomb-Scargle methods. Non-parametric and new methods have been tested as well. The modelling is implemented with Fourier series. Spline methods have also been introduced.

\item (3)~{\bf Classification} of variables in variability types under consideration. 
As the diversity of variability is very vast (see Eyer \& Mowlavi 2008), confusion of eclipsing systems with other type of variability can easily happen, in particular for contact systems. Several methods have been tested, such as Random Forest or multi-stage schemes involving Bayesian network and Gaussian mixture methods (cf. Dubath et al. 2011). 
Additional methods are being developed to improve the classification, see for example S{\"u}veges et al. 2012 for being able to distinguish eclipses from pulsation thanks to simultaneous multi-band observations.

\item (4)~{\bf Specific Object Studies}. Specific algorithms are applied to objects of given variability types as determined by Classification. The algorithms will characterize more specifically the objects knowing their variability class membership. One of the algorithms tackles eclipsing binaries, with the goal of providing a geometrical modelling of their folded light curves and subclassify them into different subclasses of eclipsing binaries.
\end{itemize}

Once the objects have been classified as eclipsing binary, they are handed over to CU4, where they will be processed through a Wilson-Devinney-type code specifically developed to automatically determine the parameters of the binary systems (see Siopis, Sadowski 2012).

Gaia is an exceptional mission for eclipsing binaries because all aspects of the mission contribute to this subject. Masses, radii and temperatures might be derived to an accuracy 
level of a few percent.

\section*{Acknowledgements} 
The authors wish to acknowledge the Coordination Unit 2 of the Gaia DPAC for the use of the GOG simulator. The simulations have been done in the supercomputer MareNostrum at Barcelona SupercomputingCenter - Centro Nacional de Supercomputaci\'{o}n (The Spanish National Supercomputing Center).
The authors would also like to thank the Gaia CU7 active members who are contributing to the Gaia pipeline.

\bibliographystyle{ceab}
\bibliography{sample}

\section*{References}
\begin{itemize}
\small
\itemsep -2pt
\itemindent -20pt
\item[] Arenou, F.: 2011, {\it American  Institute of Physics Conference Series} {\bf 1346}, 107.
\item[] Bonanos, A.~Z., Stanek, K.~Z., Udalski, A., et al.: 2004, {\it \apjl} {\bf 611}, L33.
\item[] Dubath, P., Rimoldini, L., S{\"u}veges, M., et al.: 2011, {\it \mnras} {\bf 414}, 2602.
\item[] Eyer, L., \& Bartholdi, P.: 1999, {\it \aaps} {\bf 135}, 1. 
\item[] Eyer, L., \& Mowlavi, N.: 2008, {\it Journal of Physics Conference Series} {\bf 118}, 012010.
\item[] Eyer, L., Dubath, P., Mowlavi, N., et al.: 2012, {\it IAU Symposium} {\bf 282}, 33.
\item[] Eyer, L., Palaversa, L., Mowlavi, N., et al.: 2012, {\it \apss} {\bf 341}, 207.
\item[] Mahabal, A.~A., Drake, A.~J., Djorgovski, S.~G., et al.: 2012, {\it The Astronomer's Telegram} {\bf 3872}, 1.
\item[] Mignard, F., Bailer-Jones, C., Bastian, U., et al.\ 2008, {\it IAU Symposium} {\bf 248}, 224.
\item[] Nelemans, G.: 2005, The Astrophysics of Cataclysmic Variables and Related Objects, {\it ASPC} {\bf 330}, 27.
\item[] North, P., Gauderon, R., Barblan, F., \& Royer, F.: 2010, {\it \aap} {\bf 520}, A74.
\item[] O'Mullane, W., van Leeuwen, F.: 2012, {\it Gaia Technical Note: Release scenarios for the Gaia archive} {GAIA-C9-TN-ESAC-WOM-066-02}.
\item[] Paczynski, B.: 1997, in Space Telescope Science Institute Series, The Extragalactic Distance Scale, ed. M. Livio (Cambridge: Cambridge Univ. Press), 273
\item[] Pourbaix, D.: 2011, {\it American  Institute of Physics Conference Series} {\bf 1346}, 122.
\item[] Prusti, T.: 2012, {\it Gaia Policy Document: Gaia Intermediate Data Release Scenario} {GAIA-CG-PL-ESA-TJP-011-01}.
\item[] Rauw, G., De Becker, M., Naz{\'e}, Y., et al.: 2004, {\it \aap} {\bf 420}, L9.
\item[] Richards, J.~W., Starr, D.~L., Butler, N.~R., et al.: 2011, {\it \apj} {\bf 733}, 10.
\item[] Robin, A.~C., Luri, X., Reyl{\'e}, C., et al.: 2012, {it \aap} {\bf 543}, 100.
\item[] Rucinski, S.~M.: 1996, {\it ASPC} {\bf 90}, 270.
\item[] Siopis, C., \& Sadowski, G.: 2012, Proceedings of the workshop "Orbital Couples: Pas de Deux in the Solar System and the Milky Way", eds F. Arenou \& D Hestroffer, 59 
\item[] S{\"u}veges, M., Sesar, B., V{\'a}radi, M., et al.: 2012, {\it \mnras} {\bf 424}, 2528.
\item[] Torres, G., Andersen, J., \& Gim{\'e}nez, A.\ 2010, \aapr, 18, 67 

\end{itemize}

\end{document}